\documentclass[aps,prl,twocolumn,showpacs,superscriptaddress,groupedaddress,10pt]{revtex4}
\usepackage{amsmath}
\usepackage{amssymb}
\usepackage{graphicx}
\usepackage{epsfig}
\usepackage{dcolumn}
\usepackage{bm}
\usepackage{bm}
\usepackage{blindtext}
\usepackage{natbib}
\usepackage{booktabs}

\def\be{\begin{equation}} \def\ee{\end{equation}}
\def\bea{\begin{eqnarray}} \def\eea{\end{eqnarray}}

\def\nn{\nonumber}

\begin{document}
\title{Quantum criticality of bosonic systems with the Lifshitz dispersion}
\author{Jianda Wu}
\affiliation{Department of Physics, University of California, San
Diego, California 92093, USA}
\author{Fei Zhou}
\affiliation{Department of Physics and Astronomy, University of
British Columbia, Vancouver V6T 1Z1, Canada}
\author{Congjun Wu}
\affiliation{Department of Physics, University of California, San
Diego, California 92093, USA}

\begin{abstract}
We study the quantum criticality of the Lifshitz $\varphi^4$-theory below
the upper critical dimension.
Two fixed points, one Gaussian and the other non-Gaussian, are identified
with zero and finite interaction strengths, respectively.
At zero temperature the particle density exhibits different power-law
dependences on the chemical potential in the weak and strong interaction regions.
At finite temperatures, critical behaviors in the quantum disordered
region are mainly controlled by the chemical potential.
In contrast, in the quantum critical region
critical scalings are determined by temperature.
The scaling ansatz remains valid in the strong interaction limit for
the chemical potential, correlation length, and particle density,
while it breaks down in the weak interaction one.
As approaching the upper critical dimension, physical quantities
develop logarithmic dependence on dimensionality
in the strong interaction region.
These results are applied to spin-orbit coupled bosonic systems,
leading to predictions testable by future experiments.
\end{abstract}
\pacs{73.43.Nq, 74.40.Kb, 03.75.Mn, 03.75.Nt}
\maketitle

\paragraph*{Introduction.---}
Quantum phase transitions, uniquely driven by quantum fluctuations,
appear when the ground state energy encounters non-analyticity via
tuning a non-thermal parameter.
Physical properties around quantum critical points (QCPs) are of
extensive interests because the interplay between quantum
and thermal critical fluctuations strongly influence the
dynamical and thermodynamic quantities, giving rise to
rich quantum critical properties
beyond the classical picture \cite{SpecialIssue2010,Sachdev2011a}.
Quantum critical fluctuations are believed to be responsible for
various emergent phenomena, including the non-Fermi liquid behaviors
in heavy fermion systems, unconventional superconductivity,
and novel spin dynamics in one-dimensional quantum magnets
\cite{Sachdev2011b,Gegenwart2008,Kinross2014,Wu2014}.

The progress of ultra-cold atom physics with the synthetic spin-orbit (SO)
coupling has attracted a great deal of interests
\cite{Zhouxf2013,Goldman2014,Wu2011,Stanescu2008,Gopalakrishnan2011,
Ho2011,Yip2011,LiXH2015,Lin2009,Lin2009a,Galitski2015,Zhang2016}.
In solid state systems, the SO coupled exciton condensations have also
been investigated  in semiconductor quantum wells
\cite{Wu2011,Yao2008,High2011,High2012}.
For bosons under the isotropic Rashba SO coupling, the single-particle
dispersion displays a ring minima in momentum space.
Depending on interaction symmetries, either a striped Bose-Einstein
condensation (BEC), or, a ferromagnetic condensate with a single
plane-wave, develops \cite{Wu2011,Wang2010,Gopalakrishnan2011,Ho2011,
Yip2011,Hu2012a}.
The case of the spin-independent interaction is particularly
challenging:
The striped states are selected through the ``order-from-disorder''
mechanism from the zero-point energy beyond the
Gross-Pitaevskii framework \cite{Wu2011}.
Inside harmonic traps, the skyrmion-type spin textures appear accompanied
by half-quantum vortices \cite{Wu2011,Hu2012a},
and the experimental signatures of spin textures have
already been observed \cite{High2011,High2012}.

Compared to the conventional superfluid BEC phases \cite{Weichman1986,
Weichman1988,Fisher1988,Fisher1989,Kolomeisky1992,Sachdev1994},
the progress of SO coupled bosons paves down a way to study
novel quantum criticality.
Consider an interacting Bose gas under the Rashba SO and Zeeman
couplings: When the Zeeman field is tuned to a ``critical'' value,
the dispersion minimum comes back to the origin exhibiting a
novel $q^4$-dispersion \cite{Po2014}, which is referred as
the Lifshitz-point in literature \cite{Ardonne2004}.
Quantum wavefunctions at the Lifshitz-point exhibit
conformal invariance \cite{Ardonne2004,Hsu2013,Henley1997}, which
have been applied to describe the Rokhsar-Kivelson point
\cite{Rokhsar1988} of the quantum dimer model and quantum 8-vertex model.
For the SO coupled bosons, by employing an effective non-linear
$\sigma$-model method, it is argued that at the Lifshitz point,
a quasi-long-range ordered ground state instead of a true BEC
develops due to the divergent phase fluctuations.
\cite{Po2014}.

The SO coupled bosons are not the only system to realize the Lifshitz
dispersion.
It has an intrinsic connection to a seemingly unrelated field of
quantum frustrated magnets.
Suppose a spin-$\frac{1}{2}$ antiferromagnetic Heisenberg model
defined in the square lattice with the nearest-neighbor coupling
$J_1$ and the next-nearest-neighbor coupling $J_2$.
It can be mapped to a hard-core boson model,
%with the single-particle spectrum $\varepsilon(\vec q)=-2 J_1 (\cos q_x+\cos q_%y)+4 J_2 \cos q_x \cos q_y$
and the Lifshitz dispersion appears at $J_2=J_1/2$.
These bosonic systems are fundamentally different from the regular
ones with the quadratic dispersion: They are beyond the paradigm
of the ``no-node'' theorem, or, Perron-Frobenius
theorem \cite{Feynman1972,Wu2009}, or, the Marshall-sign
rule in the context of quantum antiferromagnetism \cite{Marshall1955}.

In this article, we investigate the quantum complex $\varphi^4$-theory
with the Lifshitz dispersion below the upper critical dimension.
There exist two fixed points (FPs) -- an unstable Gaussian FP
and a non-Gaussian one with a finite interaction strength.
Quantum critical behaviors at both zero and finite temperatures
around these two FPs are investigated.
At zero temperature the particle density shows power-law
dependence on the chemical potential with different
exponents in the weak and strong interaction regions.
At finite temperatures, according to whether the chemical potential or
temperature controls the critical scalings, the disordered phase
falls into the quantum disordered or quantum critical regions, respectively.
In the quantum disordered region the power-law dependence of the chemical
potential dominates the critical behaviors, and thermal fluctuations
generate exponentially small corrections.
While in the quantum critical region, physical quantities, including
the chemical potential, correlation length, and particle density,
exhibit power-law dependence on temperature.
The scaling ansatz \cite{Sachdev2011a} breaks down in the weak
interaction limit but is sustained in the strong interaction one.
Logarithmic critical behaviors appear in both regions when the system
is near the upper critical dimension.
The connection of these results to the two-dimensional (2D) SO
coupled bosonic systems is discussed.

\paragraph*{Quantum Lifshitz $\varphi^4$-model.---}
We construct the $d$-dimensional Euclidean quantum Lifshitz
$\varphi^4$-action as
\bea
S_0  &=& T\sum \limits_{\omega _n } {\int_0^\Lambda {d^d q
\varphi ^* (-i\omega _n-\mu
+ q^4 )\varphi } } , \nonumber \\
S_I  &=& \frac{u}{2}\int_0^\beta  {d\tau \int_{1/\Lambda}^{\infty}
{d^d x\left|
{\varphi (x,\tau )} \right|^4 } },  \label{eq:lifshitz}
\eea
where $d$ is the spatial dimension;
$\Lambda$ is the ultra-violet (UV) momentum cut off;
$\mu$ and $u$ denote the chemical potential and interaction
strength, respectively; $\tau$ is the imaginary time and
$\beta=1/T$;  $\omega_n = 2n\pi T$ is the Matsubara frequency;
${\varphi (x,\tau )}$ is a complex bosonic field.
Due to the $q^4$-dispersion, the classical dimensions of $T$ and $\mu$ are
$\Lambda^4$, and that of $u$ is $\Lambda^{\varepsilon}$ where
$\varepsilon=4-d$, and thus the upper critical (spatial) dimension $d_c=4$.
In the following, we rescale $T, \mu$ and $u$ by their classical
dimensions to be dimensionless.
For quantities of the correlation length, particles density, ground state
energy that will be studied below, they are also rescaled by
$\Lambda^{-1}$, $\Lambda^d$, and $\Lambda^4$ to be dimensionless, respectively.

The zero temperature renormalization group (RG) equations are derived
following the momentum-shell Wilsonian method as presented in
the Supplemental Material (SM) \cite{SuppMat}, Sec.~A.
Two fixed points (FPs) are identified as a Gaussian FP
$(\mu^*_1,u^*_1)= (0,0)$ and a non-Gaussian one $(\mu^*_2,u^*_2)
= (0,2\varepsilon/K_d )$ appearing at $d<d_c$.
The RG equations are integrated as,
\bea
\mu_l  &=& e^{4l}\mu \label{app:epsilonmul}, \ \ \
u_l  = e^{\varepsilon l}     u/C_d(\mu,u,l),
\label{app:epsilonul}
\eea
with $l$ being the RG scale parameter.
$\mu_{l=0} = \mu$, $u_{l=0} = u$, and $C_d(\mu,u,l)=1 -
\frac{u}{8}K_d [\Phi (\mu ,1,\varepsilon /4)-e^{\varepsilon l}
\Phi (\mu_l ,1,\varepsilon /4)]$
where $K_d  = 2^{-d+1} \pi ^{-d/2} /\Gamma (\frac{d}{2})$ with $\Gamma(z)$
being the Gamma function, and $\Phi (\mu,s,\frac{\varepsilon}{4})
\equiv \sum\limits_{k = 0}^\infty{\mu^k (k + \frac{\varepsilon}{4})^{ - s}}$
the Hurwitz Lerch transcendent.
$\Phi(\mu,s,\frac{\varepsilon}{4})$ has a branch cut running from
$(+1, +\infty)$ in the complex $\mu$-plane.
Since $|\mu_l|<1$ is maintained throughout the RG process, $u_l$ remains
analytic as a function of $\mu$ .
Furthermore, in the complex $\varepsilon$-plane, $\Phi$ has
a branch cut from $(-\infty, 0)$, therefore, $\varepsilon$ can
be analytically extended to a finite positive value.

%%%%%%%%%%%%%%%%%%%%%%%%%%%%%%%%%%%%%%%%%%%%%%%%%%%%%%%%%%
\begin{figure}[t!]
\begin{center}
\includegraphics[width=7cm]{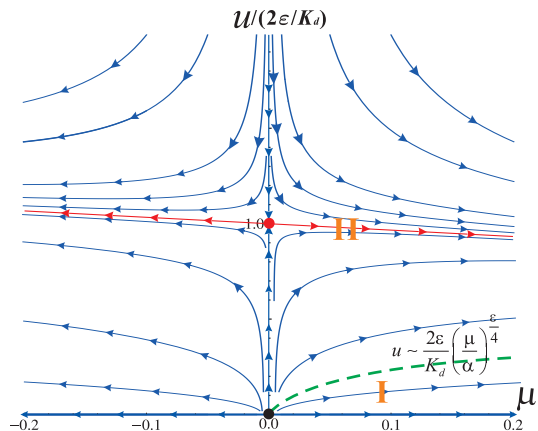}
\end{center}
\caption{Diagram of the zero temperature RG flows.
The red and black dots mark the two FPs.
Quantum phase transitions occur when $\mu$ changes sign:
The disordered and ordered phases lie at $\mu<0$ and $\mu>0$, respectively.
For $\mu>0$, symbols I and II denote the weak and strong interaction
regions, respectively.
The dashed line at $\mu>0$ marks the crossover between
these two regions.
}
\label{fig:RG_flow}
\end{figure}
%%%%%%%%%%%%%%%%%%%%%%%%%%%%%%%%%%%%%%%%%%%%%%%%%%%%%%%%%

The Gaussian FP is unstable at $\varepsilon>0$.
Close to this FP, the correlation length diverges as
$\xi(T=0,\mu)\approx
|\mu|^{-\nu}$ with the critical exponent $\nu=1/4$ rather than
$1/2$ as a consequence of the Lifshitz dispersion.
At the non-Gaussian FP, $\nu=1/4$ remains at the one-loop level
since the interaction does not renormalize the chemical potential
at zero temperature, which is different from the Wilson-Fisher FP
of the classic phase transition.

%%%%%%%%%%%%%%%%%%%%%%%%%%%%%%%%%%%%%%%%%%%%%%%%%%%%%%%
\begin{table}[h!]
\centering
\begin{tabular}{|c|c|c|c|c|}
\hline
& $u$ & $u_{l_0^* } $ & $n$ & $e_g$\\
\hline
I & $~~ u \ll \frac{{2\varepsilon }}{{K_d }}{\left(\frac{\mu}{\alpha}
\right) ^{\varepsilon /4} }~~$ & $~~{{u }}{\left(\frac{\alpha}{\mu}\right)
^{\varepsilon /4} }~~$ & $~~\mu/u~~$ & $~~\mu /2~~$\\
\hline
II& $u \gg \frac{{2\varepsilon }}{{K_d }} {\left(\frac{\mu}{\alpha}
\right) ^{\varepsilon /4} }$  & $2\varepsilon /K_d$ &
$\frac{K_d \alpha^{\varepsilon/4} }{2\varepsilon} \mu^{d/4}$ &
$\frac{\mu d}{4+d}$\\
\hline
\end{tabular}
\caption{Critical properties in the weak and strong interaction
regions.
$\mu$ is close to the phase boundary marked by $\mu=0$.
$e_g =\frac{1}{n}\int {\mu dn}$ gives the ground state energy density.
}
\label{tab:cri_zero}
\end{table}
%%%%%%%%%%%%%%%%%%%%%%%%%%%%%%%%%%%%%%%%%%%%%%%%%%%%%%%%%%

We consider the critical behaviors at zero temperature.
The RG flows based on Eq.~(\ref{app:epsilonul}) are presented
in Fig.~\ref{fig:RG_flow}.
The run-away flows indicate two stable phases: one disordered
at $\mu<0$ and the other ordered at $\mu>0$.
The disordered phase shows vanishing
particle density at the one-loop level, nevertheless,
small but finite particle density could develop beyond
one-loop at $u>0$.
The two FPs obtained above lie on the phase boundary of $\mu=0$.
To study the critical physics at $\mu>0$, a stop scale
$l_0^*$ is introduced at which $\mu_{l_0^*} =  \alpha \ll 1$.
$\alpha$ is a non-universal parameter to control the RG flow remaining
in the crossover from the critical to non-critical regions
\cite{{Sachdev1994}}.
According to different behaviors of the interaction strength $u_{l_0^*}$,
we define the weak and strong interaction regions via
$u_{l_0^*} \approx u(\frac{\alpha}{\mu})^{\frac{\varepsilon}{4}}$, or, $u_2^*$,
respectively. % as denoted by symbols I and II in Fig. \ref{fig:RG_flow}.
Correspondingly, the crossover between these two regions is approximately
marked by the line of
$u\approx \frac{2\varepsilon}{K_d}(\frac{\mu}{\alpha})^{\frac{\varepsilon}{4}}$.
The critical behaviors of the particle density $n$ and
the ground state energy density $e_g$ as well as $u_{l_0^*}$
in these two regions are summarized in
Table \ref{tab:cri_zero} [SM Sec.~A].

The finite-temperature RG equations are presented in SM Sec.~B.
We focus on two parts of the disordered region close to the QCPs:
The quantum disordered region with negative and large
chemical potential, i.e., $\mu<0$ and $|\mu|\gg T$, and the
quantum critical region with small chemical potential $|\mu|\ll T$.
Since the RG process ceases to work at $\mu_{l^*} =  - 1$, a stop scale $l^*$
is accordingly defined at which the coarse-graining length scale
reaches the correlation length $\xi(T,\mu)$.

In the quantum disordered region, the running temperature remains low
at the stop scale $l^*$, i.e.,  $T_{l^*} \ll 1$.
Similar to the zero temperature case, two different limits of the running
interaction strength are introduced, corresponding to
the weak and strong interaction regions set
by $u_{l^*} %\approx u e^{\varepsilon l^*}
\approx u \mu^{-\frac{\varepsilon}{4}}$
and $u_2^*$, respectively.
As shown in SM Sec.~C,
the correlation length is calculated as $\xi(T,\mu)
\approx |\mu|^{-1/4}\left[1 - c(T,u) \frac{T}{|\mu|}
e^{-2\frac{|\mu|}{T}}  \right] $,
where $c(T,u)=\frac{1}{8}u K_d T^{-\frac{\varepsilon}{4}}$
and $\epsilon/4$ for the weak and strong
interaction regions, respectively.
The finite temperature corrections are exponentially small.

Next consider the quantum critical region (QCR) where $T\gg |\mu|$.
Then at the stop scale $l^*$ with $\mu_{l^*}  = - 1$, $T_{l^*} \gg 1$,
indicating that the system flows into the high-temperature region.
For simplicity, we set $\mu=0$ (QCP) since in this region the correction to
thermodynamic quantities from a finite $\mu$ is sub-leading.
The correlation length, and particle density are denoted as $\xi_T$
and $n_T$, respectively.
Similarly, based on the interaction strength $u_{l^*}$
the critical behaviors at finite temperatures also fall
into weak and strong interaction regions characterized by
$u_{l^*}%\approx u e^{\varepsilon l^*}
\approx u\left[{\varepsilon}/{(2 K_d u T)}
\right]^{\frac{\varepsilon/4}{1+\varepsilon/4}}$ and $u_2^*$, respectively.
The crossover line qualitatively follows $u\sim \varepsilon T^{\varepsilon/4}$.

\paragraph*{Weak interaction region in the QCR.---}
In this region, under the condition $\ln[1/ (uT)]\gtrsim 1/\varepsilon$,
$\xi_T$ and $n_T$ are derived in SM Secs.~(D,E) as
\bea
\xi_T  \approx \left[\frac{\varepsilon}{2K_d u T} \right]^{\frac{1}{4 + \varepsilon} },
n_T  \approx a_d \left[\frac{\varepsilon }{2K_d u }
  \right]^{\frac{\varepsilon/4 }{{1 + \varepsilon/4 }}} ~T^{\frac{1}{{1 + \varepsilon/4 }}},
\ \ \,
 \label{weakden}
\eea
where $a_d  = \frac{1}{8}K_d [\psi ((4 + d)/8) - \psi (d/8)]$ with
$\psi (z) = d\ln \Gamma (z)/dz$ being the digamma function.
In this case, even though the interaction is relevant, $u_{l^*}$ remains small,
leaving a weak interaction window to justify the RG calculation.
The weak interaction
results in Eq.~(\ref{weakden}) can also be obtained following
the one-loop self-consistent method, whose details are presented
in SM Sec.~D.

The scaling ansatz is believed to be valid
for the system below the upper critical dimension \cite{Sachdev2011a}.
In our case, it dictates that the critical behavior of the correlation length in the QCR
can be cast into the form, $\xi_T \propto T^{-1/4}g({|\mu|}/{T})$,
where $g(x)$ is a universal scaling function \cite{Sachdev1994, Sachdev2011a}.
Then in the QCR, by setting $\mu=0$, the scaling ansatz predicts $\xi_T \sim T^{-1/4}$.
Nevertheless, Eq.~(\ref{weakden}) yields a novel thermal exponent for
the temperature dependence of $\xi_T$ as $\nu_T = 1/(4 + \varepsilon)$
beyond the scaling ansatz.
In contrast, typically scaling-ansatz-breakdown behaviors are
observed in systems equal to or above the upper critical dimension
\cite{Wu2011entropy,Wu2016quantum}.

When approaching the upper critical dimension $d_c$, such that
$\frac{{\Gamma (d/4,1)}}{{T^{\varepsilon /4} }} \ll \ln \left( {\frac{2}{{K_d uT}}} \right) \ll \frac{4}{\varepsilon }$,
the critical scalings
are obtained in SM. Sec.~D as
\bea
\xi_T  &\approx& \left(\frac{K_d uT }{2} \ln \frac{2}{K_d uT}\right)^{-\frac{1}{4}}, \label{weaku_marginal_corr}\\
n_T  &\approx& a_d T \left( \frac{K_d uT }{2}\ln \frac{2}{K_d uT} \right)^{-\frac{\varepsilon}{4}},
\label{weaku_marginal}
\eea
which exhibit the expected non-universal logarithmic behaviors.

Based on Eqs.~(\ref{weakden},\ref{weaku_marginal_corr},\ref{weaku_marginal}), the limits of
$u \to 0$ and $\varepsilon \to 0$
of $\xi_T$ and $n_T$ do not commute, reflecting the singular
nature of the QCP.
At finite temperatures $\xi_T$ and $n_T$ diverge as $u \to 0$ at
$\mu = 0$ (QCP), which signals the strong instability around
the unstable Gaussian FP.
Thermal fluctuations are enhanced by the Lifshitz dispersion near
the QCP due to the divergence of single-particle density of states.
Both divergences are cut off when the system has a finite $\mu$
and/or a finite interaction strength.

%%%%%%%%%%%%%%%%%%%%%%%%%%%%%%%%%%%%%%%%%%%%%%%%%%%%%%%%%%
\begin{figure}[t]
\begin{center}
\includegraphics[width=0.8\linewidth]{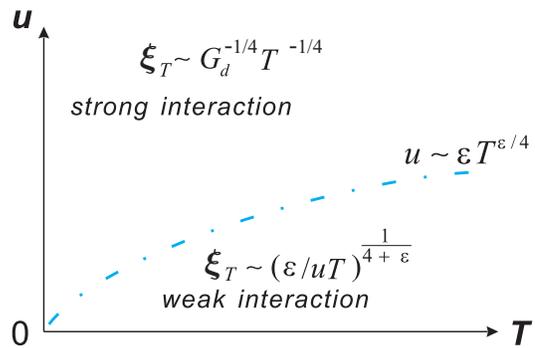}
\end{center}
\caption{A sketchy illustration for the quantum critical behaviors in
the QCR at $\mu = 0$ with a finite $\varepsilon$.
The blue dot-dashed line shows the crossover
between the weak and strong interaction regions.
}
\label{fig:crossover}
\end{figure}
%%%%%%%%%%%%%%%%%%%%%%%%%%%%%%%%%%%%%%%%%%%%%%%%%%%%

\paragraph*{Strong interaction limit in the QCR.---}
In this limit, $u \gg 2\varepsilon/ K_d$,
$\xi_T$ and $n_T$ exhibit power-law scalings as
[SM Sec.~E],
\bea
\xi_T \approx G_d^{-1/4} T^{-1/4}, \; \ \ \,
n_T \approx a_d G_d^{-1} T^{d/4},
\label{strongu3}
\eea
%then $\mu_T \approx - G_d T$ with
where $G_d  = \varepsilon \{ A +
\ln [ (1 + A\varepsilon )/(A\varepsilon ) ] \}$
and $A\approx 0.46$.
It indicates universal scaling behaviors near the non-Gaussian FP,
obeying the scaling ansatz  \cite{Sachdev2011a}.
%They are independent of the initial interaction strength as well as
%the spatial dimension $d<4$.
%The thermal exponent now becomes $\nu_T  = 1/4$.
Interestingly, at $\varepsilon \ll 1$, Eq.~(\ref{strongu3})
shows a non-analytic logarithmic dependence on $\varepsilon$ as
\be
\xi_T \approx T^{-\frac{1}{4}}
[\varepsilon \ln (1/\varepsilon )]^{-\frac{1}{4}},  \ \ \,
n_T \approx a_d T^{\frac{d}{4}}
[\varepsilon \ln (1/\varepsilon )]^{-\frac{\varepsilon}{4}}.
\label{strongu_marginal}
\ee

The above discussion for finite $\varepsilon$ in the QCR is summarized
in Fig.~\ref{fig:crossover}.
The effective interaction strength is actually temperature-dependent.
Increasing temperature enhances thermal fluctuations, which subdues
quantum fluctuations generated from the interaction.
In contrast, when decreasing temperatures, the system
gradually enters a strong interaction region
as long as $u>0$.

\paragraph*{Lifshitz Bose gas from SO coupling.---}
We apply the above general analysis to the 2D boson system with the
Lifshitz dispersion -- the SO coupled bosons under the Zeeman field.
As shown in SM Sec.~F, tuning the Zeeman field and SO coupling
strength $\lambda$ can convert the single-particle dispersion into the form,
$\varepsilon _q  =  - \mu  + \frac{{q^4 }}{{4\lambda ^2 }}$.
$\lambda$ can be used to re-scale all quantities
in the system by $ \varphi(\omega_n, \vec q) /(4\lambda ^2 )
\to \varphi (\omega_n,\vec q),\;  4 \lambda ^2 \mu \to \mu ,\;
4\lambda ^2 T\to  T , \; 4\lambda ^2 u \to  u, \; \vec q \to \vec q$.
Accordingly the low-energy physics is effectively
described by the quantum Lifshitz action Eq.~(\ref{eq:lifshitz}) at $d=2$.

%%%%%%%%%%%%%%%%%%%%%%%%%%%%%%%%%%%%%%%%%%%%%%%%%%%%%%%
\begin{table}[h!]
\centering
\begin{tabular}{|c|c|c|c|c|}
\hline
  & $u$ & $u_{l^* } $ & $n$ & $e_g$\\
    \hline
  I& $~~u \ll 8\pi\sqrt{\frac{\mu}{\alpha}}~~$ &
$~~u\sqrt{\frac{\alpha}{\mu}}~~$ & $\mu/u$ & $\mu /2$\\
    \hline
  II&   $ u \gg 8\pi \sqrt{\frac{\mu}{\alpha}}$  &
$~~8\pi ~~$
&$~~\frac{1}{8\pi }\sqrt{\alpha \mu} ~~$ &
$~~\mu/3~~ $\\
    \hline
  \end{tabular}
\caption{The zero temperature critical properties of the 2D SO
coupled bosons with the Lifshitz dispersion.
$\mu$ is close to the phase boundary. }
\label{tab:cri_zero_boson}
\end{table}
%%%%%%%%%%%%%%%%%%%%%%%%%%%%%%%%%%%%%%%%%%%%%%%%%%%%%

At zero temperature, we focus on the region at $\mu > 0$.
According to the previous analysis, when $\mu$ is close
to the phase boundary,
the crossover between the weak and strong interaction
regions is  characterized by
$u \approx 8\pi\left(\frac{\mu}{\alpha}\right)^{1/2}$.
The critical behaviors
are summarized in Table. \ref{tab:cri_zero_boson}.
%At $\mu\to 0^+$, $u_{l^*}$ is renormalized to the non-interacting limit
%in both regions since the particle density vanishes in this limit.
In the weak interaction region, $\mu=u n$ following
the mean-field result, and
in the strong interaction regime, $n \propto \sqrt{\mu}$.
In comparison, for the 2D bosons with the $q^2$-dispersion
\cite{Sachdev1994}, $n \approx \frac{\mu}{8\pi}\ln \frac{\alpha}{\mu}$
at $\mu\ll \Lambda^2$.
The relation of $n\propto\sqrt\mu$ is similar to that of 1D bosons with the
$q^2$-dispersion in the low density regime
\cite{Affleck1990,Sorenson1993,Sachdev1994}.
Such systems are well-known to be renormalized into the strong
interaction region, nearly fermionized.
This relation is also similar
to a free 2D Fermi gas with the same $q^4$-dispersion, whose single-particle
density of states also exhibits the 1D-like feature as
$\rho(\varepsilon)\propto \varepsilon^{-\frac{1}{2}}$.
Thus the dominant critical physics carries certain features of fermions.
Similar fermionization behaviors in the strongly interacting boson systems
have also been studied in the SO coupled BEC systems whose energy minima
lies in a ring in momentum space
\cite{Sedrakyan2012}, and also in the region
of resonance scattering \cite{Zhouf2013,Jiang2014}.

Similar analysis can also be applied to the ground state energy density
$e_g$.
When $u$ is sufficiently small, $e_g\approx \mu/2 \approx nu/2$ coincides
with the leading order result of the usual weak-interacting
dilute Bose gas with the $q^2$-dispersion \cite{Kolomeisky1992}.
However, in the strong interaction region, $e_g\approx \mu/3 \approx
(8 \pi n)^2/(3\alpha)$, which is very
different from  $4\pi n /[\ln{1/(4\pi n)}]$
for the case of the $q^2$-dispersion.

At finite temperatures, we focus on $\mu = 0$ in the QCR of
the 2D boson system.
The crossover between
the weak and strong interaction regions now becomes
$u \approx 2T^{1/2}$.
In the weak interaction region,
\be
\xi _T  \sim ( uT)^{ - \frac{1}{6}},  \ \ \,
n_T  \sim  u^{ - \frac{1}{3}}  T^{\frac{2}{3}},
\label{eq:weaku2dboson}
\ee
showing the divergences of $\xi_T$ and $n_T$ as $u\to 0$.
In cold atom experiments, interactions are typically weak in the
absence of Feshbach resonances, therefore, the thermal exponent
$\nu_T=1/6$ could be measurable.
Furthermore, these scaling relations
deviate from the double logarithmic behaviors of 2D boson gases with
the $q^2$-dispersion \cite{Sachdev1994}.
In contrast, in the strong interaction region,
\bea
\xi_T \sim T^{-\frac{1}{4}}, \ \ \
n_T \sim T^{\frac{1}{2}}.
\label{eq:strongu2dboson}
\eea
$\xi_T$ is nearly determined by thermal fluctuations
independent on the interaction strength.
It can be understood as a decoherent effect from the strong
inter-particle scattering.
%As a result, the quantum coherent length among particles is
%sub-leading compared with the thermal wavelength, therefore,
%to the leading order the correlation length is determined by
%the thermal fluctuations.

%Due to the Lifshitz $q^4$-dispersion, the ground state
%near the FPs might be quasi-long range ordered without
%a true BEC \cite{Ardonne2004,Po2014}.

\paragraph*{Discussion and Conclusions.---}
We have studied the quantum critical properties of a complex
$\varphi^4$-model with the Lifshitz dispersion.
At zero temperature, the particle density depends on the chemical
potential as $n\propto \mu$ and $\mu^{\frac{d}{4}}$ in the weak and
strong interaction regions controlled by the Gaussian and
non-Gaussian FPs, respectively.
At finite temperatures,  the correlation length
in the quantum disordered region scales as $|\mu|^{-\frac{1}{4}}$ in
both weak and strong interaction limits,
while the finite temperature corrections are exponentially small.
In the quantum critical region, the temperature dependence
of the correlation length scales as $\xi_T\propto T^{- \frac{1}{4+\varepsilon}}$
and $T^{-\frac{1}{4}}$ in the weak and strong interaction regions, respectively.
The critical behaviors in the weak interaction region are beyond
the scaling ansatz while it is maintained
in the strong interaction region.
In both interaction limits, logarithmic behaviors appear
when the system is close to the upper critical dimension.
The above studies based on the field-theoretical method are general,
which are applied to the 2D interacting SO coupled bosonic
system with the Lifshitz dispersion.
Their critical behaviors are testable by future experiments.

%They can also be applied to other systems such as frustrated magnetic
%systems with similar low-energy physics.

An interesting point is whether bosons with the Lifshitz dispersion
can support superfluidity.
Under the mean-field theory, the Bogoliubov phonon spectrum,
$\varepsilon_q = \sqrt{q^4 (q^4 + nu/2)}$, scales as $q^2$ in the
long wavelength limit.
It implies the vanishing of the critical velocity, and thus the
absence of the superfluidity.
In 2D, even in the ground state, the quantum depletion of the
condensate diverges signaling the possible absence of BEC even
at zero temperature \cite{Po2014}.
Nevertheless, the pairing order parameter of bosons could be non-vanishing.
It is possible that bosons at the Lifshitz-point do not
exhibit superfluidity even in the ground state with interactions,
which will be deferred for a future study.

{\it Acknowledgments} ~
We thank L. Balents for helpful discussions.
J. W and C. W. are supported by the
NSF DMR-1410375, AFOSR FA9550-14-1-0168.
F. Z. is supported by the NSERC, Canada through Discovery grant
No. 288179 and Canadian Institute for Advanced Research.
J. W. acknowledges the hospitality of Rice Center for
Quantum Materials (RCQM) where part of this work was done.

\bibliography{socview}
\bibliographystyle{apsrev-nourl}

\newpage
\appendix
%\twocolumngrid
%\setcounter{figure}{0}
%\makeatletter
%\renewcommand{\thefigure}{S\@arabic\c@figure}
%\setcounter{equation}{0} \makeatletter
%\renewcommand \theequation{S\@arabic\c@equation}

%\vskip 1.0 cm

\section{A. Zero temperature critical behaviors  of
the quantum $\varphi^4$ model with the
Lifshitz dispersion }
\label{sec:zeroT}

We start with Eq.~(1) in the main text.
Following the main text, the same rescaled dimensionless physical
variables are used.
The one-loop RG equations at zero temperature
for $d = 4 - \varepsilon$ are derived as,
\bea
\frac{{d\mu _l }}{{dl}} = 4\mu _l,  \; \ \ \,\;
\frac{{du_l }}{{dl}} = \varepsilon u_l -
\frac{u_l^2}{2} \frac{K_d}{{1 - \mu _l }},
\label{app:epsilon0TRG}
\eea
where $\mu_{l=0} =\mu $ and $u_{l=0} = u$ are the initial
chemical potential and interaction strength,
respectively. In addition, $K_d  = 2^{-d+1} \pi ^{-d/2} /\Gamma (\frac{d}{2})$
with $\Gamma(z)$ being the gamma function.
Eq.~(\ref{app:epsilon0TRG}) exhibits a Gaussian and a non-Gaussian fixed points
located at $(0, 0)$, and $(0, 2\varepsilon/K_d)$, respectively,
as shown in the main text.

When $\mu = 0$, the stop scale is infinite, {\it i.e.}, $l^*\to\infty$.
For a finite $\varepsilon=4-d>0$, the interaction $u$ is
relevant.
Following Eq.~(2) in the main text,
$u_{l \to \infty }  = 2\varepsilon /K_d  \equiv u_2^*$,
indicating flowing towards the non-Gaussian fixed point.

Now consider $\mu > 0$ but close to the FPs.
At the stop scale $l_0^*$, $\mu_{l_0^*} = \mu e^{4l_0^*} = \alpha \ll 1$,
which yields $e^{l_0^*} = (\alpha/\mu)^{1/4}$.
By integrating Eq.~(\ref{app:epsilon0TRG}) for the interaction strength,
we arrive at
\be
u_{l_0^* }
\approx \frac{{u e^{\varepsilon l_0^* } }}{{1 + K_d ue^{\varepsilon l_0^* }/(2\varepsilon) }} \approx \left\{ \begin{array}{l}
u (\frac{\alpha}{\mu})^\frac{\varepsilon}{4},\;\;\;
u \ll u_c,\\
2\varepsilon /K_d ,\;\;u \gg u_c, \\
\end{array} \right.
\label{epsilonu1qcp}
\ee
where $u_c= \frac{{2\varepsilon }}{{K_d }}
(\frac{\mu}{\alpha})^\frac{\varepsilon}{4}$.

At zero temperature,  the particle density is defined as
\bea
n = \left\langle GS|{\varphi ^* (x)\varphi (x)} |GS \right\rangle,
\eea
where $ \left\langle GS|{\cdot\cdot\cdot}|GS \right\rangle$
denotes the ground state expectation value.
The RG equation for $n$ simply follows as
\bea
\frac{dn_l}{dl} = d n_l,
\eea
with $n_{l=0} = n$ being the initial particle density,
which yields $n_{l_0^*}= e^{dl^*} n$.

At the stop scale $l_0^*$, the RG solution flows to the ordered phase,
in which the mean-field approximation \cite{Sachdev1994,Nelson1989} applies,
\bea
\mu_{l^*} = n_{l^*} u_{l^*}.
\eea
Based on Eq.~(\ref{epsilonu1qcp}), $n_{l_0^*}= e^{dl^*} n$, and
$\mu_l* = e^{4l^*} \mu$, we obtain
\bea
\mu  &=& \left( {\frac{\alpha }{\mu }} \right)^{(d - 4)/4}
\frac{{n u(\alpha /\mu )^{\varepsilon /4} }}{{1 + (\alpha /\mu )^{\varepsilon /4} K_d u/(2\varepsilon) }}\\
&=& \frac{{n u }}{{1 + (\alpha /\mu )^{\varepsilon /4} K_d u/(2\varepsilon) }}.
\eea
Consequently, the particle-density $n$ is solved as
\bea
n \approx \left\{ \begin{array}{l}
 \mu /u,\;\;\;\;\;\;\;\;\;\;\;{\kern 1pt} \;u \ll u_c , \\
 \frac{{K_d \alpha ^{\varepsilon /4} }}{{2\varepsilon }}\mu ^{d/4} ,\;\;u \gg u_c . \\
 \end{array} \right.
\eea
The average ground state energies in the weak and strong interaction
regions are expressed as
\bea
e_g =E_G /N = (1/n)\int {\mu dn} \approx \left\{ \begin{array}{l}
 \mu /2,\;\;\;u \ll u_c, \\
 \frac{{\mu d }}{{4 + d}}, \;{\kern 1pt} u \gg u_c. \\
 \end{array} \right.\label{criticaleg}
\eea

%*************************************************************
\section{B. RG equations at finite temperatures} \label{sec:cri_finite}

At finite temperatures, the RG equations are derived as
\bea
\frac{{dT_l }}{{dl}} &=& 4T_l, \\
\frac{{d\mu _l }}{{dl}} &=&
4\mu _l-\frac{{2K_d u_l }}{{e^{(1 - \mu _l )/T_l }  -1}}, \label{eq:rgmu} \\
\frac{{du_l }}{{dl}} &=& \varepsilon u_l  - K_d u_l^2 \left\{ {\frac{{\coth \left[ {\frac{{1 - \mu _l }}{{2T_l }}} \right]}}{{2(1 - \mu _l )}} + \frac{{{\rm{csch}}\left[ {\frac{{1 - \mu _l }}{{2T_l }}} \right]}}{{T_l }}} \right\},
\hspace{8mm} \label{eq:rgu}
\eea
where $T_{l=0} = T$, $\mu_{l=0} =\mu $, and $u_{l=0} = u$ are the initial
temperature, chemical potential, and interaction strength,
respectively.

The RG Eqs.~(\ref{eq:rgmu},\ref{eq:rgu}) can be formally solved as
\begin{widetext}
\bea
 T_l &=& e^{4 l}T, \\
\mu _l  &=& e^{4l} \left\{ {\mu   - 2K_d \int_0^l
{\frac{{e^{ - 4l'} u_{l'} dl'}}{{\exp \left[ {(1 - \mu _{l'} )/T_{l'} }
\right] - 1}}} } \right\} \equiv e^{4l} \mu (u,T,l), \label{runningmu}\\
 u_l  &=& e^{\varepsilon l} \left\{ {u  - K_d \int_0^l {e^{ - \varepsilon l'}
u_{l'}^2 \left[ {\frac{{\coth \left( {(1 - \mu _{l'} )/(2T_{l'} )} \right)}}{{2(1 - \mu _{l'} )}} + \frac{1}{{T_{l'} }}{\rm{csch}}^2 \left( {\frac{{1 - \mu _{l'} }}{{2T_{l'} }}} \right)} \right]} } \right\} \equiv e^{\varepsilon l} u(\mu ,T, l) \label{runningu},
 \eea
where
\bea
 \mu (u,T,l) &=&  {\mu  - 2K_d \int_0^l {\frac{{e^{ - 4l'} u_{l'} dl'}}{{\exp \left[ {(1 - \mu _{l'} )/T_{l'} } \right] - 1}}} },
 \label{renormalizemu}
 \\
 u(\mu ,T, l) &=&   {u  - K_d \int_0^l {e^{ - \varepsilon l'} u_{l'}^2 \left[ {\frac{{\coth \left( {(1 - \mu _{l'} )/(2T_{l'} )} \right)}}{{2(1 - \mu _{l'} )}} + \frac{1}{{T_{l'} }}{\rm{csch}}^2 \left( {\frac{{1 - \mu _{l'} }}{{2T_{l'} }}} \right)} \right]} },
\eea
\end{widetext}
correspond to the renormalized chemical potential and interaction strength
at the scale $l$, respectively.
These equations are the staring point to analyze the critical
behaviors in the quantum disordered and critical regions
introduced in the main text.

%%%%%%%%%%%%%%%%%%%%%%%%%%%%%%%%%%%%%%%%%%%%%%%
\section{C. Critical behaviors in the quantum disordered region}
\label{subsection:qdr}

In the quantum disordered region, $|\mu| \gg T$ and $\mu<0$, then
$T_{l^*} \ll 1$ at $\mu_{l^*} =-1$, which means the running temperature
remains small at the stop scale.
Consequently, the running interaction strength is well
approximated by its zero-temperature form,
\be
u_{l}  \approx \frac{{u e^{\varepsilon l } }}{{1 + K_d ue^{\varepsilon l }/(2\varepsilon) }} \label{eq:runningu}.
\ee
In the weak interaction limit, namely, $u \ll \frac{2\varepsilon}{K_d} e^{-\varepsilon l^*} $, the chemical potential in Eq.~(\ref{renormalizemu}) is solved as,
\be
\mu (u,T,l) \approx \mu
- \frac{{uK_d }}{2}e^{ - \left| \mu  \right|/T}
T^{1 - \varepsilon /4} T_l^{\varepsilon /4} e^{ - T_l^{ - 1} }. \label{app:disordermu}
\ee
From $\mu_{l^*} = e^{4l^*} \mu (u,T,l) = -1$, the correlation length
can be determined as,
\bea
 \xi _T  &=& e^{l^* }  \approx \left\{ {\left| \mu  \right| + TW\left( {\frac{{uK_d }}{{2T^{\varepsilon /4} e^{2\left| \mu  \right|/T} }}} \right)} \right\}^{ - \frac{1}{4}} \nonumber \\
  &\approx& \left| \mu  \right|^{ - \frac{1}{4}} \left( {1 - \frac{{\rm{1}}}{{\rm{4}}}\frac{{uK_d T^{1 - \varepsilon /4} }}{{2\left| \mu  \right|}}e^{ - 2\left| \mu  \right|/T} } \right), \label{eq:qdrweakuxi}
  \eea
where $W(z)$ is the Lambert function --- the solution of $z = W e^{W}$.
From Eq.~(\ref{eq:qdrweakuxi}), the weak-interaction condition can be
cast into
\bea
 u = \frac{2\varepsilon}{K_d} e^{-\varepsilon l^*}  = \frac{2\varepsilon}{K_d} \mu^{\varepsilon/4},
\eea
i.e., $u_{l^*} = u \mu^{-\varepsilon/4} \ll 2\varepsilon/K_d = u_2^*$.
Plugging Eq.~(\ref{eq:qdrweakuxi}) into Eq.~(\ref{app:disordermu}),
the renormalized chemical potential follows,
\be
\mu (u,T,l^*) \approx  - \left| \mu  \right| - \frac{{uK_d }}{2}T^{1 - \varepsilon /4} e^{ - 2\left| \mu  \right|/T}
\label{app:disrenweakmu}
\ee

In the strong interaction limit, namely, $u \gg 2\varepsilon/K_d = u_2^*$.
The chemical potential in Eq.~(\ref{renormalizemu}) can be calculated as
\be
\mu (u,T,l) \approx \mu  - \varepsilon Te^{ - \left| \mu  \right|/T} e^{ - T_l^{ - 1} }.
\ee
Again from $\mu_{l^*} = e^{4l^*} \mu (u,T,l) = -1$, we
determine the correlation length as
\bea
\xi _T  &=& e^{l^* }  \approx \left\{ {\left| \mu  \right|
+ TW\left( {\varepsilon e^{ - 2\left| \mu  \right|/T} } \right)} \right\}^{ - \frac{1}{4}}  \nonumber\\
  &\approx& \left| \mu  \right|^{ - \frac{1}{4}} \left( {1 - \frac{{\rm{1}}}{{\rm{4}}}\frac{{\varepsilon T}}{{\left| \mu  \right|}}e^{ - 2\left| \mu  \right|/T} } \right). \label{eq:qdrstronguxi}
\eea
Then the renormalized chemical potential in the strong interaction
region follows,
\bea
\mu (u,T, l^*) \approx  - \left| \mu  \right| -
\varepsilon Te^{ - 2\left| \mu  \right|/T}. \label{app:disrenstrongmu}
\eea
Based on Eqs.~(\ref{eq:qdrweakuxi},\ref{app:disrenweakmu},\ref{eq:qdrstronguxi},\ref{app:disrenstrongmu})
in the quantum disordered region, thermal fluctuations only give exponentially small corrections in both
weak and strong interaction regions.

%%%%%%%%%%%%%%%%%%%%%%%%%%%%%%%%%%%%%%%%%%%%%%%%%%%%
\section{D. Weak interaction limit in the QCR}
\label{subsubsection:weaku}

In the QCR with $|\mu| \ll T$, the running temperature flows into the
high-temperature region $T_{l^*} \gg 1$ at the stop scale with
$\mu_{l^*} = - 1$.
The renormalized chemical potential (Eq.~(\ref{renormalizemu})) becomes
\begin{widetext}
\be
\mu (u,T,l )\approx \mu  - \frac{{u K_d \Gamma (d/4,1)}}{2}T^{d/4} - 2u K_d T\frac{{e^{\varepsilon l}  - T^{ - \varepsilon /4} }}{\varepsilon } + O(u^2)
 \label{weaku_marginal_limit_full}
\ee
\end{widetext}
with $\Gamma (x,z) = \int_z^\infty  {t^{x - 1} e^{ - t} dt}$
being the incomplete gamma function.
Assuming $\varepsilon$ is small enough such that $\varepsilon l \ll 1$,
then the third term of Eq.~(\ref{weaku_marginal_limit_full}) becomes
$2u K_d T\left[ {l - \ln \frac{1}{{T^{1/4} }}} \right]$.
In this limit there are many different
analytic regions for the correlation length and chemical potential.
For simplicity, we focus on the region where
\bea
l \gg \ln \frac{1}{{T^{1/4} }},  \label{app:smallepsc1}
\eea
then the chemical potential in Eq.~(\ref{weaku_marginal_limit_full}) becomes
\be
\mu (u,T,l )\approx \mu  - \frac{{u K_d \Gamma (d/4,1)}}{2}T^{d/4}
- 2K_d u T l + O(u^2). \label{weaku_marginal_limit_1}
\ee
Furthermore, the third term of Eq.~(\ref{weaku_marginal_limit_1}) is asked to
dominate over the second one, which gives rise to
\bea
\frac{{4uK_d Tl}}{{uK_d \Gamma (d/4,1)T^{d/4} }}
= \frac{{4T^{\varepsilon /4} l}}{{\Gamma (d/4,1)}} \gg 1. \label{app:smallepsc2}
\eea
Under the above conditions, Eq.~(\ref{weaku_marginal_limit_full}) becomes,
\be
\mu (u,T,l) \approx \mu  - 2uK_d Tl. \label{app:weakumu}
\ee
At the stop scale $l^*$, $\mu_{l^*} = e^{4l^*} \mu (u,T,l^*) = -1$, which
gives rise to
\be
2 uK_d Tl^* e^{4l^* }  = 1 \Rightarrow l^*  \approx \frac{1}{4} \ln \frac{2}{{K_d uT}}. \label{app:smalleps}
\ee
Eq.~(\ref{app:smalleps}) automatically satisfies the condition Eq.~(\ref{app:smallepsc1})
since $\ln \left( {\frac{2}{{K_d u}}} \right)^{1/4}  \gg 1$.
Furthermore, the conditions of
$\varepsilon l^* \ll 1$ and Eq.~(\ref{app:smallepsc2}) lead to the condition
for the interaction strength,
\be
 \frac{{\Gamma (d/4,1)}}{{T^{\varepsilon /4} }} \ll \ln \left( {\frac{2}{{K_d uT}}} \right) \ll \frac{4}{\varepsilon }
\label{app:weakucondition}
\ee
%For maintaining Eq.~(\ref{app:weakucondition}), it requires
%$\frac{4}{\varepsilon } \gg \frac{{\Gamma (d/4,1)}}{{T^{\varepsilon /4} }}$,
%namely,
%\be
%\varepsilon  \ll \frac{{4T^{\varepsilon /4} }}{{\Gamma (d/4,1)}}, \label{app:smallepsc3}
%\ee
which always holds once $\varepsilon \to 0^+$ and $T \neq 0$.

Therefore, at finite temperatures as long as $\varepsilon$ is
small enough,
%Eq.~(\ref{app:smallepsc3}) is fulfilled,
the obtained stop scale
in Eq.~(\ref{app:smalleps}) self-consistently
satisfies all conditions for the analytic region we study.
From Eqs.~(\ref{app:weakumu},\ref{app:smalleps}), the renormalized chemical potential
follows,
\bea
&&\mu (u,T, l^*) \approx \mu  - \frac{K_d uT}{2} \ln \left(\frac{2}{K_d u T} \right).
\eea
Then at $\mu = 0$, the correlation length becomes,
\bea
\xi_T \approx \left[\frac{K_d uT}{2} \ln \left(\frac{2}{K_d u T} \right)\right]^{-1/4}.
\label{app:weakuxi1}
\eea

When $\varepsilon \gtrsim \ln^{-1}[1/ (uT)]$, the renormalized
chemical potential from Eq.~(\ref{weaku_marginal_limit_full}) becomes
\bea
\mu (u,T,l )&\approx& \mu  - \frac{{u K_d \Gamma (d/4,1)}}{2}T^{d/4}
- 2K_d u \frac{{Te^{\varepsilon l} }}{\varepsilon } \nonumber \\
&+& O(u^2).
\label{finalmu}
\eea
We consider the region that the third term in Eq.~(\ref{finalmu}) dominates
over the second one, which gives rise to the condition for the weak-interacting limit,
\be
\frac{{2K_d u Te^{\varepsilon l} /\varepsilon }}{{u K_d \Gamma (d/4,1)T^{d/4} /2}} \gg 1. \label{app:weaku1}
\ee
At the stop scale $l^*$, $\mu_{l^*} = e^{4l^*} \mu (u,T,l) = -1$,
then the correlation length is determined as,
\be
\xi _T  = e^{l^* }  \approx \left( {\frac{\varepsilon }{{2K_d }}} \right)^{\frac{1}{{4 + \varepsilon }}} (uT)^{ - \frac{1}{{4 + \varepsilon }}}\label{app:weaku2}.
\ee
Correspondingly, the renormalized chemical potential in Eq.~(\ref{finalmu}) follows as
\be
\mu (u,T, l^*) \approx \mu  - \left( {\frac{\varepsilon }{{2K_d }}} \right)^{ - \frac{4}{{4 + \varepsilon }}} (uT)^{\frac{4}{{4 + \varepsilon }}}.
\ee
Eqs.~(\ref{app:weaku1},\ref{app:weaku2}) lead to the condition
for the weak-interaction limit,
\be
a_1 u^{\frac{4}{\varepsilon }}  \ll T \ll 1 \label{app:weakucondition1}
\ee
where
\be
a_1  = \left[ {\frac{\varepsilon }{{2K_d }}} \right]^{ - \frac{4}{\varepsilon }} \left[ {\frac{4}{{\varepsilon \Gamma (d/4,1)}}} \right]^{ - \frac{{\varepsilon ^2 }}{{4(4 + \varepsilon )}}}.
\ee
The weak interaction condition in Eq.~(\ref{app:weakucondition1})
can be re-formulated as,
\be
u \ll \frac{\varepsilon }{{2K_d }}\left( {\frac{4}{{\varepsilon \Gamma (d/4,1)}}} \right)^{\frac{{\varepsilon ^3 }}{{16(4 + \varepsilon )}}} T^{\varepsilon /4}  \approx \frac{\varepsilon}{2K_d} T^{\varepsilon /4} \label{app:weakucondition2},
\ee
where, except the constant factor $\frac{1}{2K_d}$,
the right hand side of Eq.~(\ref{app:weakucondition2})
is just the crossover interaction strength
dividing the strong and weak
interaction regions at finite temperatures,
as silhouetted in Fig.~2 in the main text.

The above weak-interaction results can also be obtained following
the one-loop self-consistent (SC) method.
Set $\mu = 0$ (QCP), then the one-loop SC equation for the
self-energy $\mu_T^{sc}(\ll T)$ follows,
\be
\left| {\mu_T^{sc}} \right| \sim u\int_0^\infty {\frac{{q^{d - 1} dq}}{{e^{(q^4  + \left| {\mu_T^{sc}} \right|)/T}  - 1}}}  \approx \frac{{uT}}{{\varepsilon \left| {\mu_T^{sc}} \right|^{\varepsilon /4} }},
\ee
which gives rise to $\mu_T^{sc} \sim  - \left( uT/\varepsilon \right)^{\frac{1}{{1 + \varepsilon /4}}}$.
Consequently, $\xi _T^{sc} \sim (uT/\varepsilon )^{ -
\frac{1}{{4 + \varepsilon }}}$ with the same
thermal exponent as that for $\xi_T$ in Eq.~(\ref{app:weaku2}).
Furthermore, $n_T^{sc}$ becomes,
\bea
n_T^{sc} \sim \int_0^\infty  {\frac{{E^{ - \varepsilon /4} dE}}{{e^{(E + |\mu _T^{sc}|)/T}  - 1}}}  \approx \left( {\frac{\varepsilon }{u}} \right)^{\frac{{\varepsilon /4}}{{1 + \varepsilon /4}}} T^{\frac{1}{{1 + \varepsilon /4}}},
\eea
which agrees with $n_T$ in Eq.~(3) in the main text up to a constant prefactor.

%%%%%%%%%%%%%%%%%%%%%%%%%%%%%%%%%%%%%%%%%%%%%%%%
\section{E. Strong interaction limit in the QCR }
\label{subsubsection:strongu}

In the strong interaction region, $2\varepsilon /(uK_d ) \ll 1$,
{\it i.e.}, $u \gg 2\varepsilon/K_d$.
When reaching the stop scale $l^*$, we determine the
renormalized chemical potential (Eq.~(\ref{renormalizemu})) as follows,
\bea
\mu (u,T, l) &\approx& \mu  - A\varepsilon T - \varepsilon T\ln \left[ {\frac{{e^{4l} T(1 + A\varepsilon )}}{{1 + A\varepsilon e^{4l} T}}} \right] \nonumber \\
&+&O(\varepsilon^2),
\label{strongu_full}
\eea
with $A = \ln [e/(e-1)] \approx 0.46$.
Therefore at $\varepsilon \ll 1$, since $T_l = T e^{4 l} \gg 1$
in the QCR, the third term in Eq.~(\ref{strongu_full}) dominates
over the second term.
In this case, at $\mu = 0$, $\mu_{l^*} = e^{4l^*} \mu (u,T,l) = -1$ leads to
\be
\varepsilon Te^{4l^* } \ln \left[ {\frac{{e^{4l^* }
T(1 + A\varepsilon )}}{{1 + A\varepsilon e^{4l^* } T}}} \right] = 1,
\ee
which is solved as
\be
Te^{4l^* }  =  - \frac{1}{{\varepsilon \left( {A - W[e^A (A + (1/\varepsilon ))]} \right)}}.
\ee
Expanding the Lambert function as
\begin{widetext}
\be
W[e^A (A + (1/\varepsilon ))] = \ln \frac{{e^A }}{\varepsilon } - \ln \ln \frac{{e^A }}{\varepsilon } + \frac{{\ln \ln \frac{{e^A }}{\varepsilon }}}{{\ln \frac{{e^A }}{\varepsilon }}} - \frac{{\ln \ln \frac{{e^A }}{\varepsilon } - \frac{1}{2}\ln ^2 \ln \frac{{e^A }}{\varepsilon }}}{{\ln ^2 \frac{{e^A }}{\varepsilon }}} + O(\varepsilon ) \approx A - \ln \varepsilon,
\ee
we arrive at
\bea
Te^{4l^* }  =  - \frac{1}{{\varepsilon \left( {A - W[e^A (A + (1/\varepsilon ))]} \right)}} \approx \frac{1}{{\varepsilon \ln (1/\varepsilon )}} \Rightarrow \xi _T  = e^{l^* }  \approx \left[ {\varepsilon T\ln (1/\varepsilon )} \right]^{ - 1/4}.
\eea
Therefore at $\varepsilon \ll 1$,
\be
\mu (u,T, l^*) \approx \mu  - \varepsilon T[\ln \left( {1/\varepsilon } \right) +\ln \ln \left( {1/\varepsilon } \right)] \approx \mu  - \varepsilon T\ln \left( {1/\varepsilon } \right) = \mu  + \varepsilon T\ln \varepsilon \;\rm{when\;\varepsilon \ll 1}.
\ee

At finite $\varepsilon$, the third term in Eq.~(\ref{strongu_full})
is comparable with the second one, then
\be
\mu (u,T, l^*)  = \mu  - A\varepsilon T - \varepsilon T\ln \left[ {\frac{{1 + A\varepsilon }}{{\frac{1}{{e^{4l} T}} + A\varepsilon }}} \right] \approx \mu  - A\varepsilon T - \varepsilon T\ln \left[ {\frac{{1 + A\varepsilon }}{{A\varepsilon }}} \right] = \mu  - G_d T,
\ee
\end{widetext}
where $G_d  = \varepsilon \left\{ {A +
\ln \left[ {(1 + A\varepsilon )/(A\varepsilon )} \right]} \right\}$.
From the stop-scale condition $\mu_l^* = -1$, we reach ($\mu = 0$)
\be
\xi_T  = e^{l^* }  \approx G_d^{ - 1/4} T^{ - 1/4}. \label{app:strongxi}
\ee

The particle density in the QCR at the stop scale $l^*$ can be derived as
\bea
n_T  &=& K_d e^{ - dl^* } \int_0^1 {\frac{{q^{d - 1} dq}}
{{e^{\left( {q^4  + 1} \right)/T_{l^* } }  - 1}}} \nonumber \\
&\approx& K_d T_{l^* } e^{ - dl^* } \int_0^1 {\frac{{q^{d - 1} dq}}{{q^4  + 1}}}
 = a_d Te^{\varepsilon l^* }, \label{app:particledensity}
\eea
where $e^{l^*} = |\mu(u,T, l^*)|^{-1/4}$ and
\be
a_d  = \frac{{K_d }}{8}\left[ {\psi \left( {\frac{{4 + d}}{8}} \right)
- \psi \left( {\frac{d}{8}} \right)} \right]
\ee
with $\psi (z) = d\ln \Gamma (z)/dz$ and $\Gamma(z)$ are digamma and
gamma functions, respectively. Since in the quantum disordered region
$\xi_T = e^{l^*}$ is finite, thus Eq.~(\ref{app:particledensity})
indicates the particle density vanishes at zero temperature. Nevertheless,
a small particle density could appear when RG calculation is carried
out beyond one loop.
Plugging the correlation lengths of
Eqs.~(\ref{app:weakuxi1},\ref{app:weaku2},\ref{app:strongxi})
into Eq.~(\ref{app:particledensity}),
particle densities in the QCR under different situations
are derived as presented in Eqs.~(3,5,6) in the main text.

%%%%%%%%%%%%%%%%%%%%%%%%%%%%%%%%%%%%%%%%%%%%%%%%%%%%%%%%%%%%%%%
\section{F. Derivation of Lifshitz-type Action from the 2D Bose gas }
\label{sec:derivation}

We consider the following Hamiltonian $H = H_0 + H_I$ defined as
\bea
H_0  &=&\int d^2\vec q ~\psi^\dagger_\alpha(\vec q)
h_{\alpha\beta}(\vec q) \psi_\beta(\vec q), \label{H0} \\
H_I  &=& \frac{u}{2}\int d^2\vec r~
\psi^\dagger_\alpha(\vec r) \psi^\dagger_\beta(\vec r)
\psi_\beta(\vec r) \psi_\alpha(\vec r),
\label{HI}
\eea
where $\vec q = (q_x, q_y)$ and
$h(\vec q)= - \mu  + \frac{1}{2m}[q_x^2  + q_y^2 -
2\lambda (\sigma^x q_x  + \sigma^y q_y )+ 2\lambda \Omega \sigma^z ]$.
In Eqs.~(\ref{H0},\ref{HI}), $\psi_\alpha$ is the bosonic annihilation
operator; the pseudospin indices $\alpha, \beta = \uparrow, \downarrow$
refer to two different internal components;
$\sigma^{\mu}$'s are the Pauli matrices associated with the spin
components $S^{\mu} = \frac{1}{2}\sigma^{\mu}$ ($\mu = x,y,z$);
$\lambda$ and $\lambda \Omega$, reduced by $2m$, are the isotropic
Rashba SO strength and Zeeman coupling, respectively;
$u$ is the $s$-wave scattering interaction.
Eqs.~(\ref{H0},\ref{HI}) describe a two-dimensional interacting
Bose gas with an isotropic Rashba spin-orbit coupling
under a Zeeman field.
The quadratic part, $H_0$, yields the single-particle spectra
of two branches as $\varepsilon _{q_{\pm} }  =
- \mu + ( q^2  \pm 2\lambda \sqrt {\Omega ^2  + q^2 } )/(2m)
$ with $q = |\vec q|$.

We work in the regime of a large Zeeman splitting field and large
Rashba SO coupling strength, therefore, for the low energy physics,
only the lower branch of $\varepsilon_{q_-}$ is considered.
The global minimum of $\varepsilon_{q_-}$ is either located at $q = 0$ if
$\lambda<\Omega$, or, at $q = \sqrt{\lambda^2 - \Omega^2}$ if
$\lambda \geq \Omega$.
At $\lambda = \Omega$, the two minima merge into one with
a quartic low-energy dispersion as $\varepsilon_{q_-}=-\mu+ q^4/(8m\lambda^2)$
(the minimum energy reference point $-\lambda^2/m$ is shifted to zero),
where large $\lambda$  implies that the band given
by $\varepsilon_{q_-}$ is almost flat.

An effective action for the low-energy bosons is constructed as follows.
The Rashba SO coupling is assumed strong enough such that only the lower
branch bosons needs to be considered.
We assume that bosons are almost fully polarized with the Zeeman
field at small values of $\vec q$, and thus the Berry phase effect
associated with the variation of spin eigenstates with $\vec q$ neglected.
The boson field variable is denoted as $\varphi(\vec x,\tau)$
with the momentum cut-off  defined as $\Lambda$ inversely proportional to
the average interaction range in real space.
Following the method of bosonic coherent state path integral
\cite{Negele1998},
we write down the low-energy effective action $S = S_G + S_I$ with
the quartic single-particle dispersion at $\lambda=\Omega$
in the imaginary time formalism as
\bea
S_G  &=&T \sum_{\omega_n}\int_0^\Lambda
{d^2 \vec q ~\varphi ^ *  (\omega_n, \vec q)\left[{-i\omega_n -\mu +
\frac{q^4}{4\lambda^2}} \right]\varphi(\omega_n, \vec q)}, \label{actiong}
\nn \\
S_I  &=& \frac{u}{2} \int_0^\beta d\tau \int_{1/\Lambda} {d^2 \vec x\left|
{\varphi (\vec x ,\tau )} \right|^4 },  \label{app:actioni}
\eea
where
%$\varphi$ is the bosonic field corresponding to the lower branch;
%$\omega_n=2n\pi T$ is the bosonic Matsubara frequency;
$2m$ is absorbed into $\lambda$.
%The momentum cut-off is defined as $\Lambda$ inversely proportional to
%the average interaction range in real space.
The powers of $\Lambda$ can be used as the natural units of different physical quantities.
The units of $T$, $\omega_n$, and $\mu$ are $\Lambda^2$,
and those of $\lambda$ and $\varphi(x,\tau)$ are $\Lambda$.
$u$ is dimensionless.

For the situation we are interested in, $\lambda$ is always finite,
which now can be used to re-scale all quantities in
Eqs.~(\ref{app:actioni}) by $ \varphi(\omega_n, \vec q) /(4\lambda ^2 )
\to \varphi (\omega_n,\vec q),\;  4 \lambda ^2 \mu \to \mu ,\;
4\lambda ^2 T\to  T , \; 4\lambda ^2 u \to  u, \; \vec q \to \vec q$.
Then the action of Eqs.~(\ref{app:actioni}) is converted to the action
of Eq.~(1) at $d=2$ in the main text.
Following the analysis in the main text, at zero temperature two FPs
are immediately identified as $(\mu^*_1/\Lambda^4,u^*_1/\Lambda^2)= (0,0)$
and $(\mu^*_2/\Lambda^4,u^*_2/\Lambda^2) = (0,8\pi)$.

\end{document}